\documentclass[letterpaper]{sig-alternate-2013}
\usepackage{booktabs}
\usepackage{enumitem} 
\usepackage[caption=false]{subfig}
\usepackage{array,multirow}
\usepackage{xcolor}
\usepackage{times}
\usepackage{bm}
\usepackage{lastpage}
\usepackage[keeplastbox]{flushend}
\usepackage{pgfplots}
\usepackage{color, colortbl}
\usepackage[hidelinks]{hyperref}
\usetikzlibrary{patterns}

\newcommand{\descr}[1]{\smallskip\noindent\textbf{#1}}

\makeatletter
\def\url@leostyle{%
  \@ifundefined{selectfont}{\def\UrlFont{}}%
  {\def\UrlFont{}}%
} \makeatother 
\renewcommand{\footnotesize}{\scriptsize}

\newenvironment{customlegend}[1][]{%
    \begingroup
    \csname pgfplots@init@cleared@structures\endcsname
    \pgfplotsset{#1}%
}{%
    \csname pgfplots@createlegend\endcsname
    \endgroup
}%

\def\addlegendimage{\csname pgfplots@addlegendimage\endcsname}

\makeatletter
\def\@copyrightspace{\relax}
\makeatother

\makeatletter
\def\@copyrightspace{\relax}
\makeatother
\begin{document} 

\pagenumbering{arabic}

\title{Kissing Cuisines: Exploring Worldwide\\Culinary Habits on the Web$^{\thanks{A preliminary version of this paper appears in WWW 2017 Web Science Track.}}$}
\author{Sina Sajadmanesh$^{\star}$, Sina Jafarzadeh$^{\star}$, Seyed Ali Osia$^{\star}$, Hamid R. Rabiee$^{\star}$, Hamed Haddadi$^{\dagger}$\\[0.5ex]
Yelena Mejova$^{\ddagger}$, Mirco Musolesi$^{\sharp}$, Emiliano De Cristofaro$^{\sharp}$, Gianluca Stringhini$^{\sharp}$\\[1ex]
\affaddr{$^{\star}$Sharif University of Technology, $^{\dagger}$Queen Mary University of London}\\
\affaddr{$^{\ddagger}$Qatar Computing Research Institute, $^{\sharp}$University College London}\\[1ex]
}

\maketitle

\begin{abstract}
Food and nutrition occupy an increasingly prevalent space on the web, and dishes and recipes shared online provide an invaluable mirror into culinary cultures and attitudes around the world. More specifically, ingredients, flavors, and nutrition information become strong signals of the taste preferences of individuals and civilizations. However, there is little understanding of these palate varieties. 
In this paper, we present a large-scale study of recipes published on the web and their content, aiming to understand cuisines and culinary habits around the world. Using a database of more than 157K recipes from over 200 different cuisines, we analyze ingredients, flavors, and nutritional values which distinguish dishes from different regions, and use this knowledge to assess the predictability of recipes from different cuisines. We then use country health statistics to understand the relation between these factors and health indicators of different nations, such as obesity, diabetes, migration, and health expenditure. Our results confirm the strong effects of geographical and cultural similarities on recipes, health indicators, and culinary preferences across the globe.

 \end{abstract}

\section{Introduction} 
\label{sec:intro}

Nowadays, food has become an essential part of today's digital sphere and an important source for our social media footprints. 
New jargon has entered our vocabulary with expressions like ``foodie'', ``food porn'', and ``food tourism'', hint at the buzz around the entertainment arising from our culinary experiences.
With the rise of social media, and the proliferation of always-on always-connected devices, this gobbling revolution is not 
confined to our kitchens, restaurants, and food stalls, but naturally breaks out on the social web.
Sharing pictures of one's food has become a growing passion for both tourists and locals~\cite{mejova2016fetishizing},
and dedicated food searching and sharing apps, along with recipe websites and the ubiquitous social presence of celebrity chefs, have all contributed to a thriving culture and passion around food worldwide. 

Around the world, different cuisines are naturally intertwined with cultures, traditions, passions, and religion of individuals living in different countries and continents. Sushi, curry, kebab, pasta, tacos -- these are just examples of foods conventionally associated with specific countries, as are specific cuisines and ingredients. Different dietary habits around the world are also closely related to various health statistics, including cancer incidence~\cite{armstrong1975environmental}, death rates~\cite{keys1986diet}, cardiovascular complications~\cite{michel1999mediterranean},  and obesity~\cite{kratz2013relationship}. 

Although there are many common beliefs about cuisines, recipes, and their ingredients, it is still unclear what types of ingredients are unique in/about different countries, what factors make cuisines similar to each other (e.g., in terms of ingredients or flavors), and how these factors are related to individuals' health. With this motivation in mind, in this paper, we set to investigate the way in which ingredients relate to different cuisines and recipes, as well as the geographic and health significances thereof. We use a few datasets, including 157K recipes from over 200 cuisines crawled from Yummly, BBC Food data, and country health statistics. 

\descr{Overview \& Contributions.} First, we characterize different cuisines around the world by their ingredients and flavors. Then, we train a Support Vector Machine classifier and use deep learning models to predict a cuisine from its ingredients. This also enables us to discover the similarity across different cuisines based on their ingredients -- e.g., Chinese and Japanese -- while, intuitively, they might be considered different. We look at the diversity of ingredients in recipes from different countries and compare them to geographic and human migration statistics. We also measure the relationship between the nutrition value of the recipes vis-\`a-vis public health statistics such as obesity and diabetes. 

\descr{Paper Organization.} The rest of the paper is organized as follows. In Section~\ref{sec:data}, we present the datasets used in our study, then Section~\ref{sec:ingredients} presents an analysis of the diversity of the ingredients around the world, looking at geographic diversity patterns of cuisines and notable ingredients in particular ones. In Section~\ref{sec:simil}, we look at the similarity between the cuisines based on their ingredients and flavors, and use these results to train machine-learning classifiers for ingredient-based cuisine prediction models in Section~\ref{prediction}. In Section \ref{sec:nutrition}, we correlate the nutrition values of recipes for different countries with their public health statistics. After reviewing related work in Section~\ref{sec:related}, the paper concludes in Section~\ref{sec:conclusions}.

\section{Datasets}
\label{sec:data}
Our study relies on a number of datasets, namely, a large set of recipes collected from Yummly, a list of ingredients compiled by BBC Food, and country health statistics. In this section, we describe these datasets in detail.

\subsection{Yummly data}\label{sec:dataset:yummly}

Yummly is a website offering recipe recommendations based on the user's
taste.\footnote{\url{http://www.yummly.com}} It allows users to search for
recipes, learning which dishes the user likes and providing them with recipe suggestions. 
It also provides a user-friendly API, which we use to collect recipes. 
First, we crawled Wikipedia for a list of
cuisines\footnote{\url{https://en.wikipedia.org/wiki/List\_of\_cuisines}}, then, in Summer 2016, 
we queried the Yummly API for recipes belonging to each cuisine.
In the end, we obtained 157,013 recipes belonging to over 200 different cuisines. 
Due to API restrictions, we limited the number of recipes per to 5,000. %

Each recipe obtained from the Yummly API contains a number of attributes. In our
study, we use the  following:

\setlist[enumerate]{itemsep=0mm}
\begin{enumerate}
\item \textbf{Ingredients:} Each recipe contains a list of the ingredients that
  are required to prepare it. Since Yummly acts as a recipe aggregator from
  various cooking sites, the ingredients do not always appear with the same
  wording. In fact, it is very common to see the same ingredient written
  with different spellings or by using a different terminology. We overcome
  these issues through a standardization process described in Section~\ref{sec:dataset:bbc}.

\item \textbf{Flavors:} Recipes are identified by six flavors, specifically, saltiness, sourness, sweetness, bitterness, savoriness, and spiciness. These scores are on a range of 0 to 1.

\item \textbf{Rating:} Users are encouraged to provide a rating, from 1 to 5,  for the recipes
  that they try. We use the average review rating for each recipe as a measure of its popularity. 

\item \textbf{Nutrition:} Unfortunately, the Yummly search API does not directly provide 
 nutritional information for the recipes. As a consequence, we designed a simple
  web crawler to fetch the corresponding web page for each recipe in our
  dataset, and extract information on the amount of protein, fat, saturated fat,
  sodium, fiber, sugar, and carbohydrate of a recipe (per serving), as well as calories.
\end{enumerate}

Although some ingredients appear in other languages (e.g., German, French, etc), the recipes presented here are mostly in English; hence it is possible that some more authentic or niche local recipes might be missing from our dataset. However, considering the number of recipes and a large cut-off threshold introduced later on, we are confident this does not significantly affect our analysis.
Moreover, authors of the recipes might not represent the entire population, given the fact that they are likely to be tech-savvy. This might introduce a potential bias in the dataset, but at the same time, this potential issue is compensated by its richness in terms of the variety of dishes from different countries available in it.

\subsection{BBC Food Data}\label{sec:dataset:bbc}

BBC Food\footnote{\url{http://www.bbc.co.uk/food/}} is a part of the BBC website providing information about recipes, ingredients, chefs, cuisines, and other information related to cooking and dishes from all BBC programs. %
In Summer 2016, we crawled all the ingredients from the BBC Food website, collecting about 1,000 ingredients, which we used to organize and standardize the ingredients in the Yummly dataset. The standardization process is as follows:

\setlist[enumerate]{itemsep=0mm}
\begin{enumerate}[label=(\roman*)]
\item We extracted all the 11,000 ingredients from the Yummly dataset and performed a preliminary data cleaning, i.e., removing measurement units (mass, volume, etc), numbers, punctuation marks, and other symbols.

\item Due to the multilingualism of the Yummly data, we used the Google Translate API to perform automatic language detection and translation of all the Yummly ingredients to English.

\item We used the BBC list of ingredients as a reference, and mapped all possible ingredients from the Yummly list to it.

\item As not all ingredients from the Yummly list were successfully mapped, we merged the similar ones into groups, and the ingredients in each group were manually mapped to its representative ingredient.
\end{enumerate}
Overall, this process yields about 3,000 standardized ingredients.

\subsection{Country health statistics}

As diet is directly related to the health of individuals, we also set to relate Yummly statistics to real-world health data. 
To this end, we will use the diabetes prevalence estimates from World Development Indicators by The World Bank\footnote{\url{http://data.worldbank.org/indicator/SH.STA.DIAB.ZS}}, the health expenditure as a percentage of total GDP from The World Bank\footnote{\url{http://data.worldbank.org/indicator/SH.XPD.TOTL.ZS}}, and the obesity prevalence from the World Health Organization\footnote{\url{http://apps.who.int/gho/data/view.main.2450A}} in the countries to which the cuisines are mapped, using the most recent available data, which is from 2014.

\begin{figure*}[!t]
\centering
\subfloat[Global diversity]{
\includegraphics[width=0.475\textwidth]{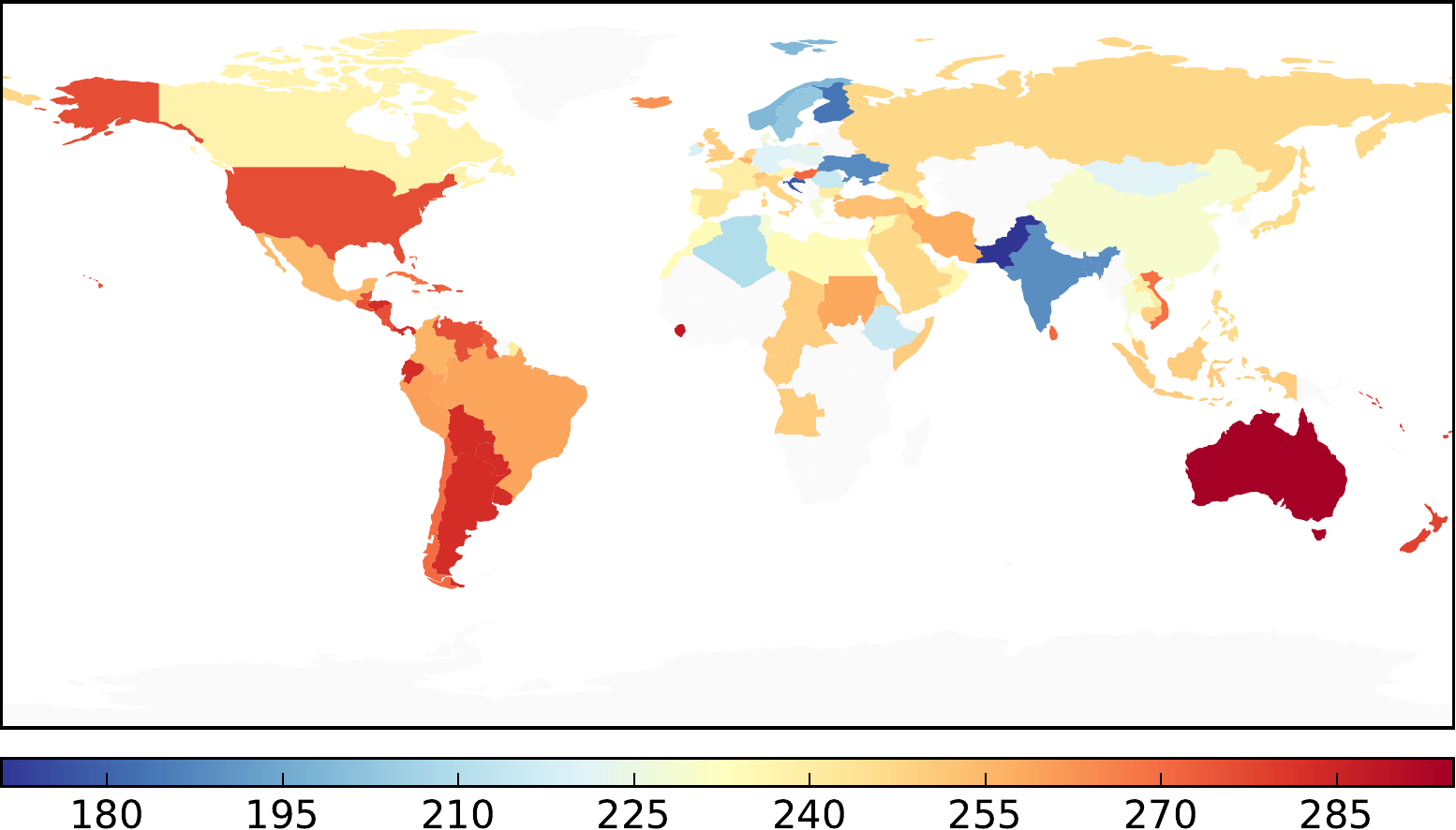}
\label{fig:div:global}
}
\hfil
\subfloat[Local diversity]{
\includegraphics[width=0.475\textwidth]{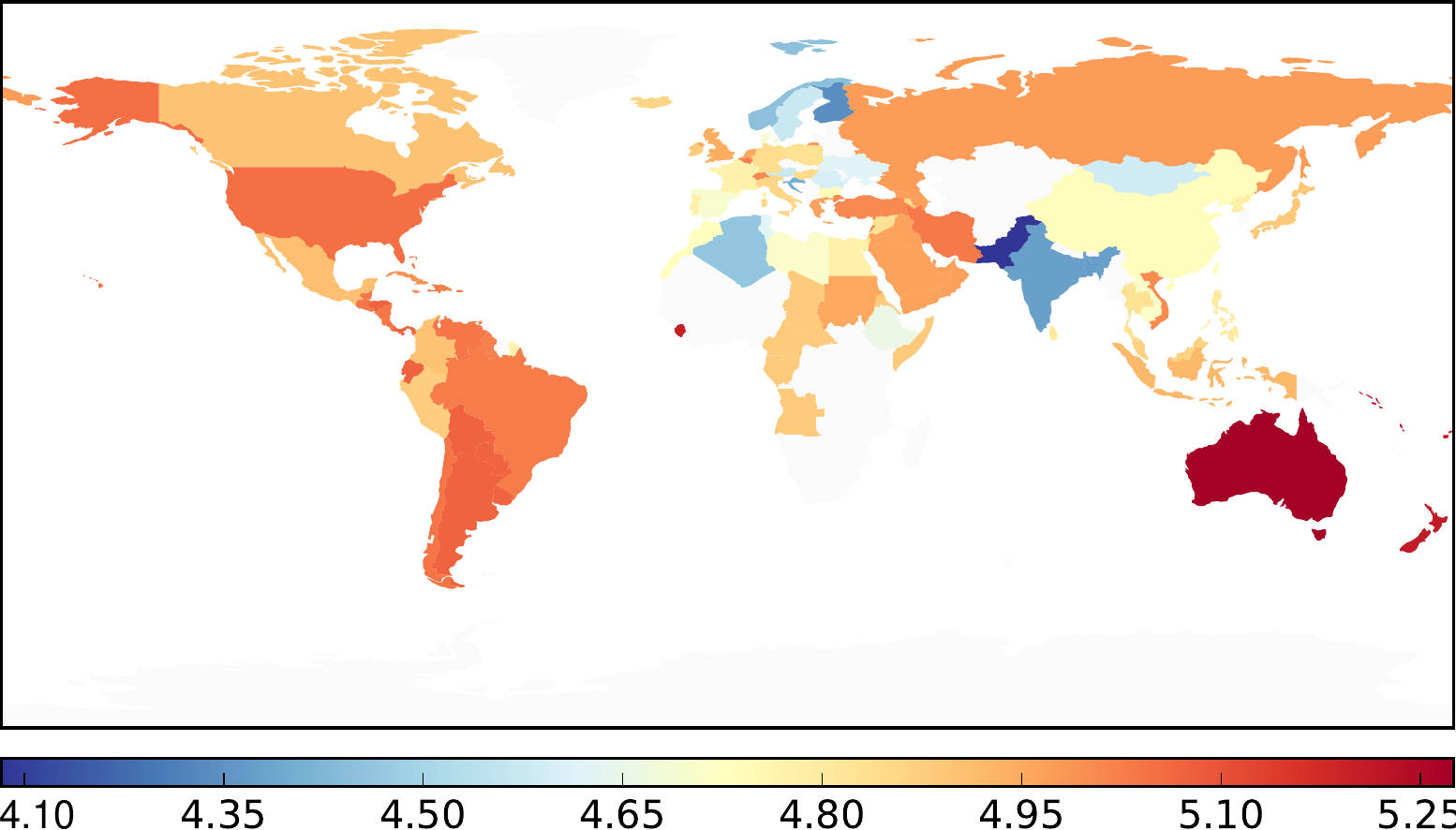}
\label{fig:div:local}
}
\caption{Diversity of ingredients used in dishes around the world. The dark blue reflects the least diverse countries while the dark red shows the most diverse ones.}
\label{fig:div}
\end{figure*}

\section{Ingredients Around the World}
\label{sec:ingredients}

In this section, we provide a characterization of the ingredients used in dishes from all over the world. First, we investigate the diversity of ingredients in different countries. Next, we define the concept of ``complexity'' of a dish in terms of its ingredients and look at how complexity changes around the world. Finally, we discuss a series of case studies of most notable and significant ingredients in some eminent cuisines.

\subsection{Diversity of ingredients}\label{sec:ingredients:diversity}

Aiming to investigate the diversity of ingredients in dishes of a cuisine, we set to answer the following questions:

\setlist[enumerate]{itemsep=0mm}
\begin{enumerate}
\item How many different unique ingredients are used in total in dishes of each country? In other words, what is the number of unique ingredients the people of a country have ever used to prepare a culinary dish? The answer to this question is what we refer to as the \textit{global diversity}.
\item How different are the dishes of an individual country relative together in terms of their ingredients combination? In other words, do different dishes usually share some ingredients or their ingredients are almost different? The answer to this question is what we call \textit{local diversity}.
\end{enumerate}

The local and global diversity of ingredients in a country depend on many parameters including the geographical location, climatic conditions, agricultural situation, or even the amount of immigration which directly influences the diversity of culinary cultures. The calculation of the global diversity is performed in two steps. Since the number of recipes per different cuisines are variable, we first set a fixed number of 100 recipes per cuisine, discarding cuisines containing fewer number of recipes, and sampling from cuisines containing more number of recipes uniformly at random, to have an equal number of recipes in all cuisines. This results in a final set of 82 different cuisines each containing 100 recipes. We then map the result obtained for each cuisine to its corresponding country. Some countries are mapped with more than one cuisine, for these, we record the average result over their associated cuisines.

To calculate the local diversity, we look at each cuisine as a probability distribution over all standard ingredients. By counting the total number of occurrences of each ingredient in all recipes of a particular cuisine, and then normalizing the values such that they sum to one, we obtain the ingredient distribution for that cuisine. We then calculate the entropy of these distributions as the local diversity of their corresponding cuisines. The entropy of the ingredient distribution measures the unpredictability of ingredients used in the dishes. Therefore, the higher the entropy of the ingredient distribution of a particular cuisine, the more different the ingredients combination of its recipes, and thus the higher the local diversity. To preserve the smoothness of the ingredient distributions, we again keep the 82 cuisines with more than 100 recipes. After calculating the local diversity for each cuisine, we follow the same procedure as for the global diversity to map the cuisine-based results to countries.

Figure~\ref{fig:div} shows the local and global diversities of ingredients for different countries around the world. The local and global diversities have a meaningful correlation with each other. The countries with high global diversity have also high local diversity, and countries with low global diversity tend to have low local diversity as well. This happens because as the global diversity increases, people will have more options to choose as the ingredients for their foods, so they can prepare relatively different dishes.

\begin{figure}[t!]
\centering
\begin{tikzpicture}[trim axis left, trim axis right]
\begin{axis}
[
small,
width=0.85\columnwidth,
xmax=40000,
ymin=160,ymax=300,
ylabel=Global Diversity,
xlabel=Net Migration,
x label style={at={(axis description cs:0.5,0.05)},anchor=north},
y label style={at={(axis description cs:.1,.5)},anchor=south},
xmajorgrids,
ymajorgrids,
max space between ticks=20pt,
try min ticks=5,
]
\addplot[color=purple,ultra thick] table{figs/global-migration.txt};
\end{axis}
\end{tikzpicture}
\caption{Relationship between the global ingredient diversity and the average net migration of different countries}
\label{fig:div:mig}
\end{figure}
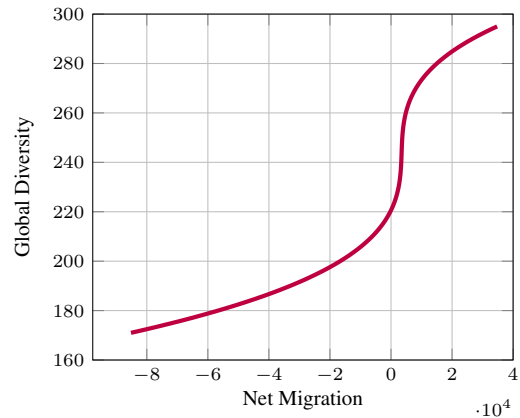

Another interesting trend from Figure~\ref{fig:div} is that countries like the United States and Australia, which usually accept a high number of immigrants, have a relatively high ingredient diversity. Regarding this, we hypothesized that the number of immigrants coming to a country must have an influence on the ingredient diversity of that country. To investigate this fact, we collected the net migration data from the World Bank\footnote{\url{http://data.worldbank.org/indicator/SM.POP.NETM}} which shows the difference between the total number of immigrants and emigrants during a time period. We correlated the global diversity with the average net migration from 1960 to 2016. To this end, we fitted a polynomial curve to the data points considering the global diversity and the net migration of 100 countries. The result is illustrated in Figure \ref{fig:div:mig}. As expected, an increase in the net migration results in an increase in the global diversity of ingredients. When the net migration is above zero, for which the countries accept more immigrants than emigrants, the increase in global diversity is much more considerable compared to the case having negative net migration. This is mainly due to immigrants bringing their native culinary culture with themselves, which in turn makes the cuisines of their target country richer.

\newcommand{\complexityplot}{
\begin{tikzpicture}[trim axis left, trim axis right]
\begin{axis}
[
small,
width=0.85\columnwidth,
legend pos=south east,
ymin=0,ymax=1,
xmin=0.0,xmax=35,
xmajorgrids,
ymajorgrids,
xlabel=\# ingredients,
ylabel=Cumulative Probability,
x label style={at={(axis description cs:0.5,0.05)},anchor=north},
y label style={at={(axis description cs:.1,.5)},anchor=south},
y tick label style={
    /pgf/number format/.cd,
        fixed,
        fixed zerofill,
        precision=1,
    /tikz/.cd
},
xtick={0,5,...,40},
legend entries={Norwegian, Tunisian, Lao},
legend style={font=\scriptsize},
]
\addplot[color=purple,ultra thick] table{figs/Norwegian.txt};
\addplot[color=orange,ultra thick,densely dotted] table{figs/Tunisian.txt};
\addplot[color=cyan,dashed,ultra thick] table{figs/Lao.txt};
\end{axis}
\end{tikzpicture}
}

\begin{figure}[!t]
\centering
\complexityplot
\caption{Cumulative complexity distribution of dishes for some representative cuisines.}
\label{fig:complexity:cdf}
\end{figure}

\begin{figure}[!t]
\centering
\includegraphics[width=0.99\columnwidth]{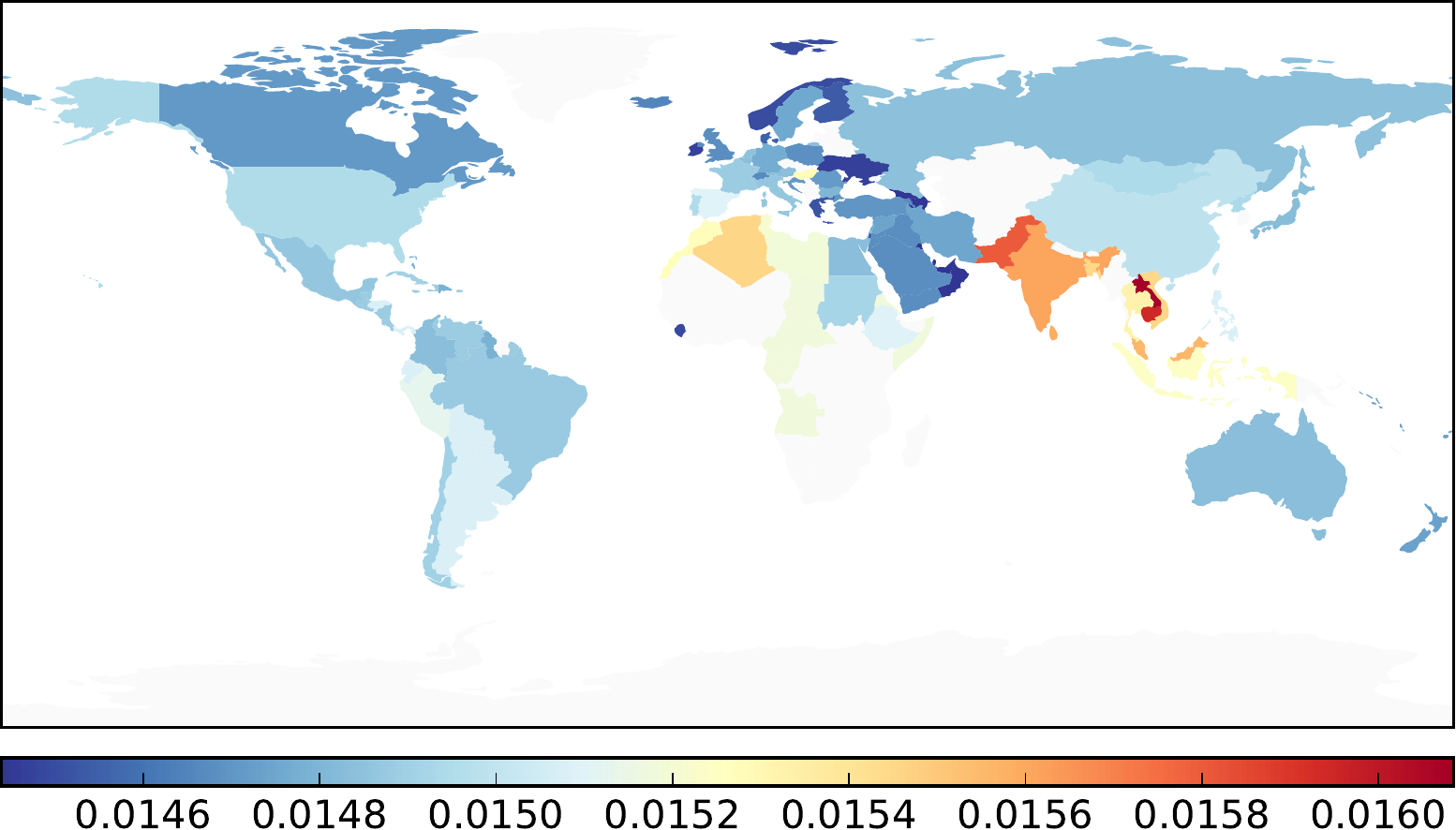}
\caption{Complexity of dishes around the world. Countries with the least complex dishes are shown in dark blue while the ones with most complex plates are depicted in dark red.}
\label{fig:complexity}
\end{figure}

\begin{figure*}[!t]
\centering

\subfloat[Italian]{
\includegraphics[width=0.3\textwidth]{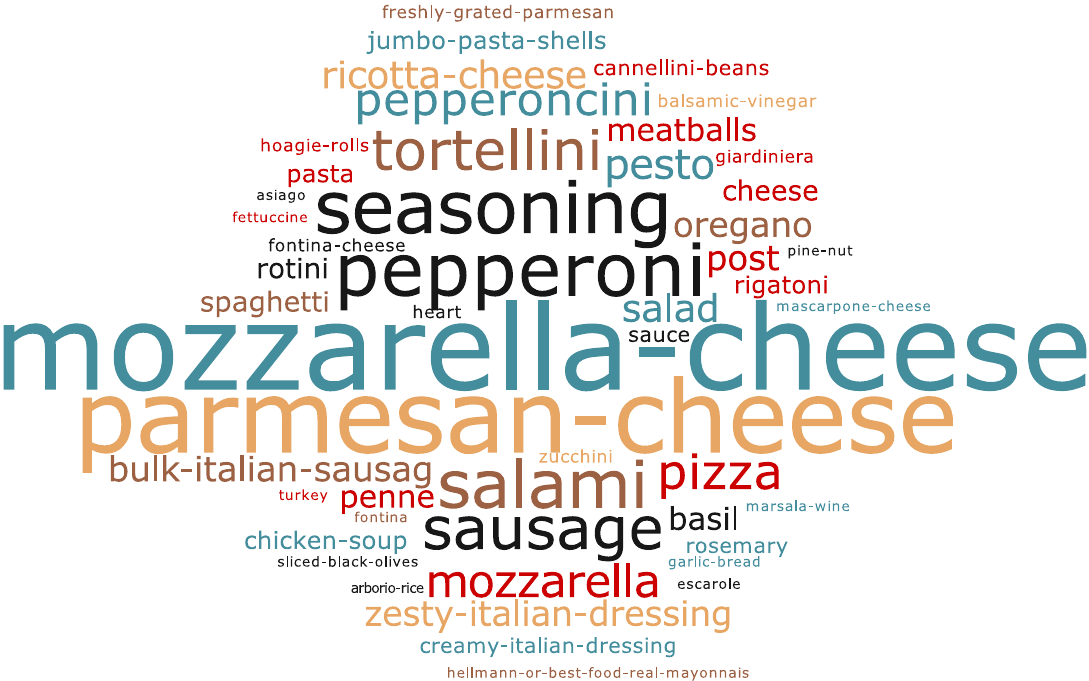}
\label{fig:not:it}}
\hfil
\subfloat[Indian]{
\includegraphics[width=0.3\textwidth]{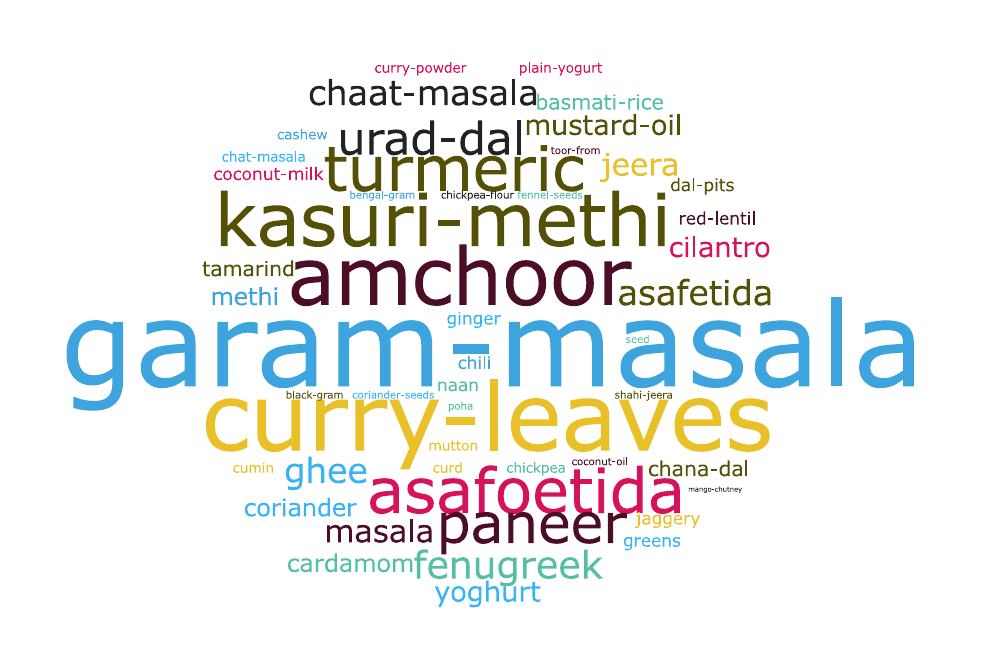}
\label{fig:not:in}}
\hfil
\subfloat[Mexican]{
\includegraphics[width=0.27\textwidth]{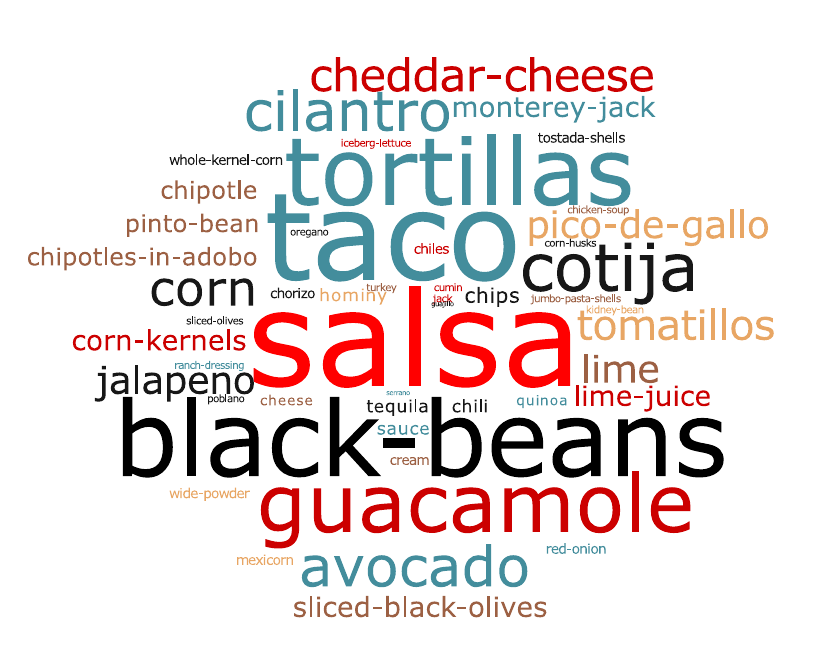}
\label{fig:not:mx}}

\caption{Notable ingredients in Italian, Indian, and Mexican cuisines. More notable ingredients have been drawn larger.}
\label{fig:not}
\end{figure*}

\subsection{Complexity of dishes}

Another interesting concept about the culinary preferences of different countries is the complexity of dishes. The complexity of a dish is simply the number of unique ingredients required to prepare it. Accordingly, a cuisine is more complex than another one if its dishes are proportionally more complex than the another's.

Formally speaking, each cuisine is associated with the complexity distribution of its dishes. For a sample cuisine, this distribution, namely $P(X=i)$, specifies the probability of a dish from that cuisine to have exactly $i$ unique ingredients. This way, the cumulative complexity distribution (CCD) will give us an insight about the complexity of dishes in a particular cuisine.

Figure~\ref{fig:complexity:cdf} depicts the cumulative complexity distribution (CCD) for Norwegian, Tunisian, and Lao cuisines as an illustrative example. We observe that the CCD for Norwegian cuisine grows faster than the others, while for Lao, it is relatively slower. As a result, about half of the Lao dishes have more than 15 ingredients, while for Norwegian cuisine, this fraction is below 10\%. This means that Lao dishes are relatively more complex than Norwegian ones. Thus for each cuisine, the area under its CCD is inversely related to its complexity. Hence, we use the reciprocal of the area under CCD as a measure of complexity for a cuisine.

Figure \ref{fig:complexity} shows the complexity of dishes for different countries around the world. Here we have used the same approach as in Section \ref{sec:ingredients:diversity} to map the cuisines to countries. Except for some cases, the complexities are consistent with the diversities. This is due to the fact that as the number of available ingredients increases in a country (which is the result of global diversity,) people can leverage more ingredients and prepare more complex dishes. The exceptions here are China and India, two countries with the most population in the world. the complexity of dishes in these countries are relatively high, while their ingredients diversity is low. 
This can be the result of overpopulation or special culinary culture in these countries. Perhaps, these countries had or have good chefs that could cook more complex foods with the available ingredients!

\subsection{Notable ingredients} \label{sec:ingredients:notable}

Due to the geographical locality of the ingredients, specific cuisines are mostly associated with different sets of ingredients. Some of these ingredients are used worldwide, while there are some others which are local to specific cuisines. We call the latter kind of ingredients ``notable'' since they tend to signify the cuisines in which they are used.
We now study the most notable ingredients associated to some well-known cuisines using our dataset of recipes. An ingredient is more notable to a specific cuisine if (1) It is used in most dishes of that cuisine; and (2) It is barely used in dished of other cuisines.

We use the Term Frequency - Inverse Document Frequency (TF-IDF) to find notable ingredients in each cuisine. In this approach, each ingredient is considered as an atomic word, and the collection of all the ingredients appeared within a cuisine is considered as a document. A TF-IDF calculation leads us to find the weight of each ingredient in the corpus of documents. This way, we can specify the importance of each ingredient within each cuisine.

Figure~\ref{fig:not} shows the top-50 most notable ingredients for Italian, Indian, and Mexican cuisines as a case study. We also looked at other similar cases, which we do not present here due to space constraints. The bigger the name of an ingredient, the more distinctive it is in its associated cuisine. The soundness of results can be easily verified using Google Trends.\footnote{\url{https://www.google.com/trends}} For example, the term ``Mozzarella'' has the highest search frequency in Italy, while ``Garam masala'' is the most popular food additive in India according to its search volume.
\section{Similarity of Cuisines}
\label{sec:simil}

In this section, we set to determine the similarities between cuisines, using a number of different methods and data.

\subsection{Ingredient-based similarity}

At first, we calculate the similarity between different cuisines based on the ingredients used in their recipes. To this end, we convert cuisines into vector space, representing each cuisine as a vector where each element indicates the frequency of an specific ingredient in that cuisine. Thereby, for each cuisine we obtain an ingredient-based feature vector which we leverage to calculate the similarity between different cuisines.
If we normalize each ingredient-based feature vector such that the elements of a vector sum to one, then each vector will represent a probability distribution over standard ingredients. This way, we can use the distance measures proposed for probability distributions as a measure of similarity between two vectors. For this purpose, we use \textbf{Jensen-Shannon (JS) divergence}, which is defined between two probability distributions $P$ and $Q$ as:
\begin{displaymath}
JS(P,Q)=\frac{1}{2}\left[KL(P\parallel M)+KL(Q\parallel M)\right]
\end{displaymath}
where $M=\frac{1}{2}(P+Q)$ and $KL(P\parallel M)$ is the Kullback-Leibler (KL) divergence from $M$ to $P$. Since the JS divergence is a distance measure between 0 and 1, we take $1-JS(P,Q)$ as the similarity measure between two cuisines with their associated ingredient distributions $P$ and $Q$. We have used JS divergence instead of the simpler KL divergence because $KL(P\parallel Q)$ goes to infinity when for an ingredient like $i$, $P(i)$ is non-zero while $Q(i)$ is. This case almost always happens in our data due to the geographical locality of ingredients. Therefore, we turned to JS divergence which does not have this drawback.

\begin{figure*}[t!]
\centering
\subfloat{
\definecolor{NA}{RGB}{255, 0, 0}
\definecolor{LA}{RGB}{255, 106, 0}
\definecolor{WE}{RGB}{182, 255, 0}
\definecolor{EE}{RGB}{0, 255, 33}
\definecolor{AF}{RGB}{64, 144, 144}
\definecolor{ME}{RGB}{0, 148, 255}
\definecolor{SA}{RGB}{0, 38, 255}
\definecolor{EA}{RGB}{178, 0, 255}
\definecolor{OC}{RGB}{255, 0, 110}
\begin{tikzpicture}
\tiny
\begin{customlegend}[
legend columns=9,
legend style={
draw=none,
column sep=1ex,
},legend entries={North America,Latin America,Western Europe,Eastern Europe,Middle East,Africa,South Asia,East Asia,Oceania}]
\addlegendimage{NA,fill=NA,area legend}
\addlegendimage{LA,fill=LA,area legend}
\addlegendimage{WE,fill=WE,area legend}
\addlegendimage{EE,fill=EE,area legend}
\addlegendimage{ME,fill=ME,area legend}
\addlegendimage{AF,fill=AF,area legend}
\addlegendimage{SA,fill=SA,area legend}
\addlegendimage{EA,fill=EA,area legend}
\addlegendimage{OC,fill=OC,area legend}
\end{customlegend}
\end{tikzpicture}
\label{fig:sim:leg}}
\setcounter{subfigure}{0}

\subfloat[Ingredient-based similarity]{
\includegraphics[width=0.45\textwidth]{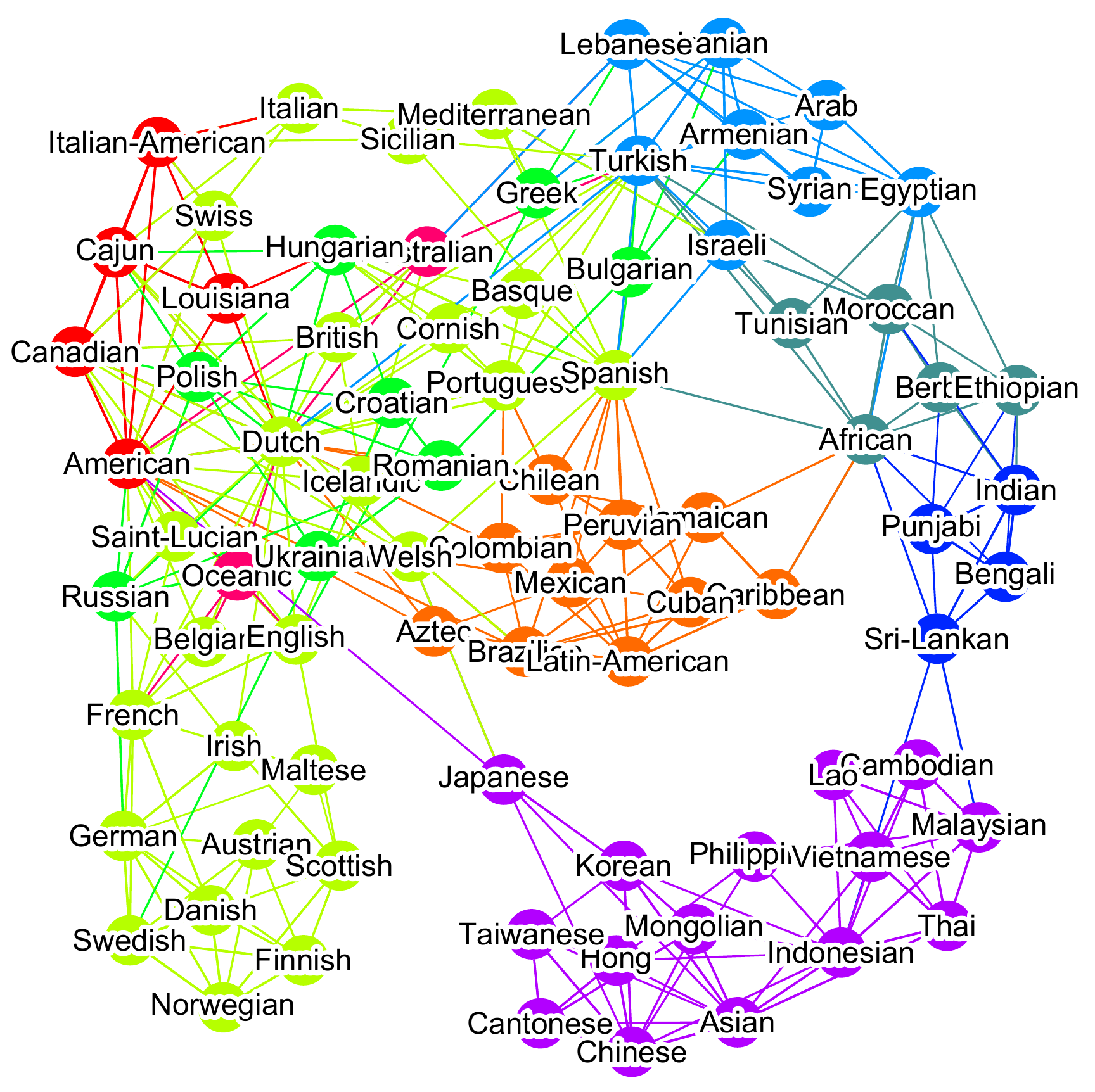}
\label{fig:sim:ing}}
\hfil
\subfloat[Flavor-based similarity]{
\includegraphics[width=0.45\textwidth]{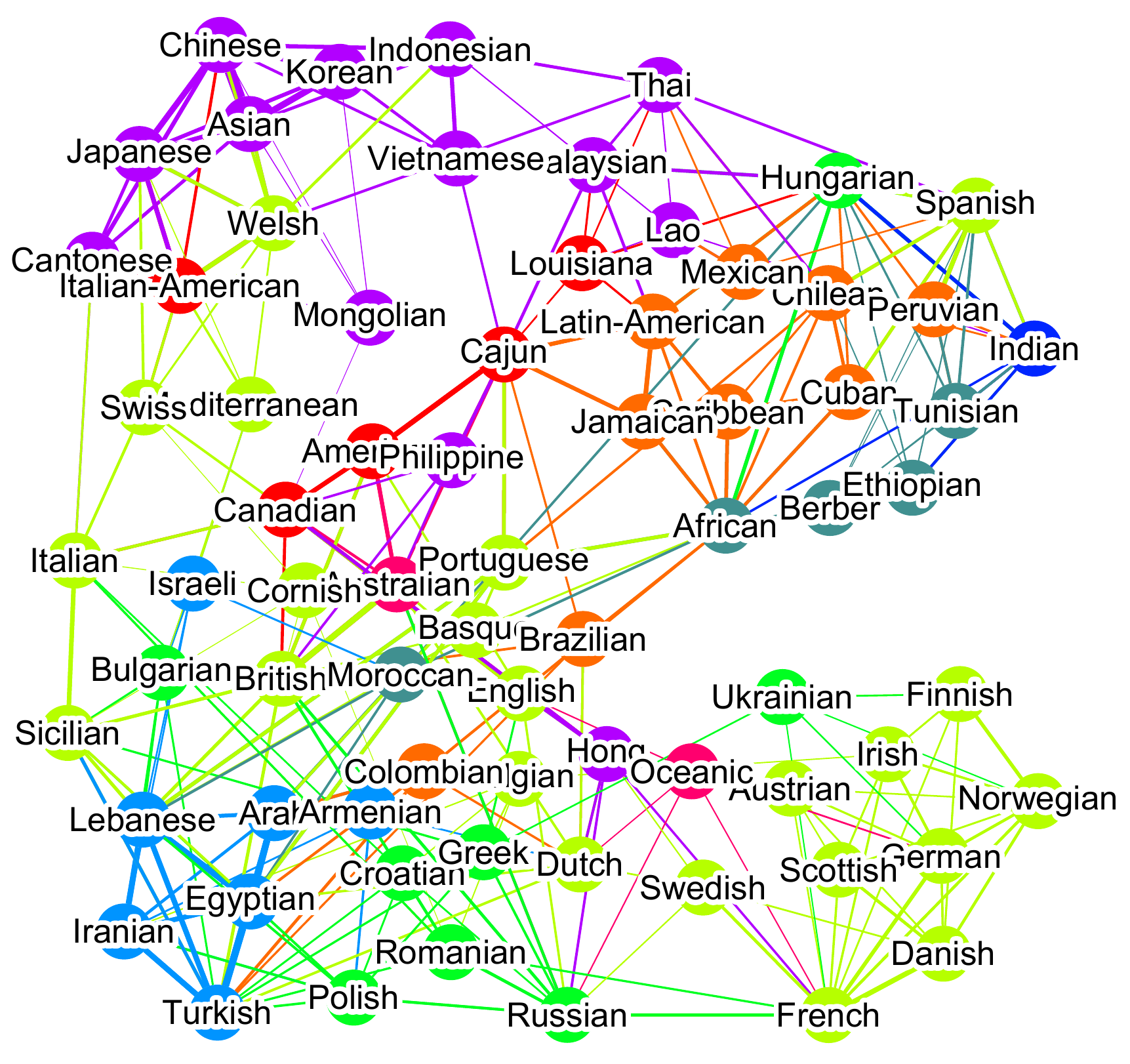}
\label{fig:sim:flv}}

\caption{Graph of similarity between different cuisines in terms of their ingredients and flavors. Each cuisine is linked with five most similar ones. Color of a cuisines denote the geographical region it resides in.}
\label{fig:sim}
\end{figure*}

Using the above similarity metric, we calculated all similarities between each pair of cuisines. To assert the smoothness of ingredient distributions needed to compute JS divergence, we limited our cuisines to those 82 ones having more than 100 recipes. Figure \ref{fig:sim:ing} illustrates the obtained results in a graph-based fashion. In this graph, each node represents a cuisine and is linked to its top-5 most similar cuisines. Link weights are proportional to the obtained similarity score between two endpoints. We colored each cuisine node according to the geographical region it resides in, including North America, Latin America, Africa, Western Europe, Eastern Europe, Middle East, South Asia, East Asia, and Oceania. To visualize the graph, we have used \textit{ForceAtlas} graph drawing algorithm implemented in Gephi tool. This a force-directed algorithm which makes densely connected nodes to be grouped together \cite{Noack2009, waltman2010unified} and thus the communities become revealed.

Figure \ref{fig:sim:ing} shows that cuisines which reside in the same region are more similar to themselves and thus have been grouped together. For example, we clearly see the clusters formed by Eastern and Southern Asian, Middle Eastern and African, Latin American, and Western European cuisines. This indicates that the geography has a direct impact on the ingredients people use for their dishes. Furthermore, due to the similarity of cultures in Europe and North America, and even Oceania, it can be seen that clusters formed by the cuisines of these regions greatly overlap with each other. Therefore, ethnicity and culture can also greatly affect the culinary habits of the people.

There are examples here that confirm the impact of migrations on the culinary culture of a region. For instance, since most of the South American countries were former colonies of either Spain or Portugal, we see that Spanish and Portuguese cuisines are both very close to the Latin American ones. The same holds for the Oceanic and North American cuisines, where their corresponding countries were formerly the colonies of the United Kingdom, being similar to British cuisine. However, as opposed to the Latin American cuisines, the similarity of Oceanic and American cuisines are not confined to British cuisine, due to the high rate of immigrations and thus the diversity of the population.

\subsection{Flavor-based similarity}

In addition to the ingredient-based similarity, we calculate the similarity between cuisines in terms of the flavors provided in their recipes. This can help us understand how different cuisines are related to each other based on the taste of their dishes.

As mentioned in Section~\ref{sec:dataset:yummly}, each recipe contains the flavor scores for six different flavors including saltiness, sourness, sweetness, bitterness, savoriness, and spiciness. To calculate the similarity between cuisines based on these flavors, as done for ingredient-based similarity, we consider each cuisine as a distribution over different flavors. As different flavors of a recipe are correlated to each other -- for instance, a dish can hardly be both sweet and spicy simultaneously -- and due to the continuity of flavor scores, we hypothesize that the flavor scores are sampled from a multivariate Gaussian distribution, where each covariate corresponds a particular flavor. Considering this assumption, we fit a multivariate Gaussian distribution to each cuisine so that each one becomes associated by a mean vector representing the average of flavor scores over all of its recipes, and a covariance matrix representing how flavors change relative to each other within that cuisine. 

After fitting a multivariate Gaussian distribution to each cuisine using maximum likelihood estimation, we use KL divergence to measure the distance between the distributions associated to each pair of cuisines. As KL divergence is an asymmetric measure, for each pair of cuisines with $P$ and $Q$ as their corresponding flavor distributions, we use $\left[\frac{1}{2}(KL(P\parallel Q)+KL(Q\parallel P))\right]^{-1}$ as a symmetric similarity measure between them.

Figure \ref{fig:sim:flv} shows the result of flavor-based similarity between different cuisines in a graph-based manner. We followed exactly the same steps as in Figure \ref{fig:sim:ing} to draw the graph, except that we used flavor-based similarity between cuisines. We observe that even though the flavors are not as much discriminant as ingredients, still we can observe some geographical patterns. For instance the clusters formed by Eastern Asian, Middle Eastern, Latin American, and Northern European cuisines are clear in this case as well. But what is obvious here is the fact that although there is a sense of taste similarity between the dishes from neighboring countries, the flavors are naturally shared all over the world.

Generally, some of the similarities depicted in Figure \ref{fig:sim} appear to be fallacious at first look, but after precise inspections, we find them to be valid. For example, the Welsh cuisine is found to be similar to Asian cuisines, which seems to be somewhat peculiar. By further investigation, we have found out that Asians are the second major ethnic group in Wales, based on the 2011 census\footnote{\url{http://www.ons.gov.uk/ons/dcp171778\_290982.pdf}}. Therefore, it seems that the Welsh cuisine is mostly influenced by the Asian migrants and their culinary culture. Another interesting example is Indian cuisine, which is found to be similar to African and Ethiopian cuisines. We have found that both of the African and Ethiopian cuisines mostly contain spicy dishes, like Indian cuisine which is famous for its spicy plates. Accordingly, they share many spices like ginger, cardamom, cinnamon, chili pepper, and clove, which in turn makes them to be similar to each other. This can be verified by referring to the Wikipedia page of these cuisines\footnote{See \url{https://en.wikipedia.org/wiki/Ethiopian\_cuisine\#Traditional\_ingredients} and \url{https://en.wikipedia.org/wiki/North\_African\_cuisine}}.

\newcommand{\barplot}[1]{
\begin{tikzpicture}
\begin{axis}
[
ybar,
small,
width=0.6\columnwidth,
enlarge x limits=0.4,
legend style={at={(0.5,1.3),font=\footnotesize},
anchor=north,legend columns=-1, draw=none},
symbolic x coords={Acc, F1},
xtick=data,
ymin=0.4,
ymax=0.75,
ymajorgrids,
ytick={0.4,0.45,...,0.8},
y tick label style={
/pgf/number format/.cd,
fixed,
fixed zerofill,
precision=2,
/tikz/.cd
},
]
\addplot+[postaction={
    pattern=crosshatch dots,
	pattern color=red,
}] table[x=Measure,y=Value]{figs/svm-#1.txt};
\addplot+ table[x=Measure,y=Value,]{figs/dnn-#1.txt};
\legend{SVM, DNN}
\end{axis}
\end{tikzpicture}
}

\section{Cuisine Classification}\label{prediction}

We now address the question of ``\textit{How good we can predict a recipe's cuisine, given its ingredients?}''. The answer to this question can help us understand the fact that how good a combination of ingredients can represent a cuisine, as opposed to Section \ref{sec:ingredients:notable} where ingredients were singly considered as cuisines signatures. To answer this question, We use two different classifiers, Support Vector Machine (SVM), which is previously used in \cite{su2014automatic} for the same task, and Deep Neural Network (DNN), which is popular nowadays for classification purposes. To extract a feature vector for each recipe, we convert it into a boolean bag of words vector, considering each ingredient as an atomic word. Therefore, each recipe is represented as a vector with a length equal to the total number of ingredients, which is 3,286. The labeling of recipes are performed according to one of the following settings:
\begin{itemize}
\item \textbf{Cuisine Prediction:} Each recipe is labeled to its cuisine. we consider 82 different cuisines having more than 100 recipes as different classes, resulting in about $100K$ recipes.\\
\item \textbf{Region Prediction:} Each recipe is labeled according to one of the 9 geographical regions where its cuisine belongs to. The regions are considered the same as in Section~\ref{sec:simil}. This results to have about $157K$ recipes.
\end{itemize}

For multi-class classification with SVM, we use linear kernel with one vs. rest coding. The class imbalance problem is resolved with adjusting the weight of each cuisine inversely proportional to its frequency. The implementation is done using Scikit-learn machine learning library in python \cite{sklearn_api}. For DNN, we use Keras deep learning library \cite{chollet2015keras} and create four dense hidden layers and a softmax output layer. Each of the first two hidden layers consists of 1000 neurons, and the two last ones each have 500 neurons. Dropout regularization \cite{srivastava2014dropout} is used for all of the hidden layers. We use Adadelta \cite{zeiler2012adadelta} with default parameters as the optimizer. For both methods we take $80\%$ of the data as training set and the remaining $20\%$ as the test set. The prediction performance of both methods are evaluated under accuracy and F-measure.

\begin{figure}[t!]
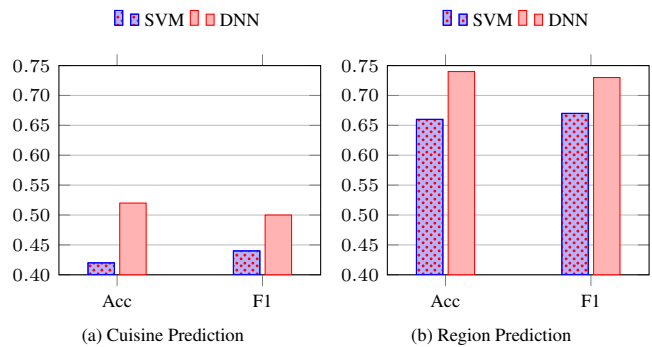

\centering
\subfloat[Cuisine Prediction\label{fig:prediction:cuisine}]{\barplot{cuisine}}
\subfloat[Region Prediction\label{fig:prediction:region}]{\barplot{region}}
\caption{The prediction performance of different methods for cuisine and region prediction tasks.}
\label{fig:prediction}
\vspace{-0.2cm}
\end{figure}

Figure~\ref{fig:prediction} shows the results with both SVM and DNN, Figure \ref{fig:prediction:cuisine} illustrates those for cuisine prediction, while Figure \ref{fig:prediction:region} the region prediction task. The DNN model performs about $24\%$ better than SVM for cuisine prediction task under accuracy and over $13\%$ better under F-measure. For region prediction task, since the number of classes are much fewer than cuisine prediction, both methods performed relatively better. In this case, the accuracy and F-measure achieved by the DNN model is about $12\%$ and $9\%$ better relative to those achieved by SVM, respectively.

Aiming to shed light on the similarity of recipes in different regions, we use the confusion matrix of the DNN model for region predictions in Table~\ref{table:conf}. Each region name is abbreviated in two letters, e.g., LA denotes Latin American and AF African cuisines. The number of correctly classified recipes are shown in bold and for each class, the greatest number of miss-classifications is shown in red. This table clearly demonstrates that almost all of the miss-classifications fall under Western European. This is probably due to the huge ethnic composition of Western European countries which resulted in the diversity of culinary cultures of that region. The table shows that for some regions like Southern and Eastern Asian, the number of miss-classified recipes are somewhat low relative to the correctly classified ones. This result is analogous to Figure \ref{fig:sim:ing} in which these regions were almost disconnected from the others. On the other hand, for some cuisines like Oceanic, Eastern European, and Northern America, the number of miss-classifications are relatively high, mostly with Western European. This is due to the fact that the cultures in these regions are very similar to each other, mainly due to the common ethnics and history.

\begin{table}
\centering
\caption{Confusion Matrix for DNN Region Prediction}
\label{table:conf}
\scriptsize
\begin{tabular}{l l c c c c c c c c c}
\toprule
& & \multicolumn{9}{c}{{\footnotesize Prediction Outcome}} \\
\cmidrule(l){3-11}
& & LA & SA & OC & EA & AF & WE & ME & EE & NA\\
\midrule %
\multirow{9}{*}{\rotatebox{90}{{\footnotesize Actual Class}}}
& LA & \textbf{1888} & 15 & 2 & 84 & 25 & \color{red}455 & 13 & 33 & 92\\
& SA & 18 & \textbf{961} & 1 & \color{red}52 & 16 & 40 & 17 & 5 & 3\\
& OC & 21 & 2 & \textbf{177} & 21 & 3 & \color{red}119 & 5 & 6 & 18\\
& EA & 49 & 37 & 2 & \textbf{5211} & 13 & \color{red}342 & 26 & 24 & 51\\
& AF & 31 & 23 & 2 & 28 & \textbf{704} & \color{red}136 & 57 & 7 & 15\\
& WE & 453 & 49 & 21 & \color{red}660 & 85 & \textbf{9430} & 165 & 541 & 557\\
& ME & 35 & 24 & 9 & 107 & 72 & \color{red}366 & \textbf{634} & 78 & 17\\
& EE & 51 & 30 & 1 & 58 & 29 & \color{red}885 & 41 & \textbf{1320} & 94\\
& NA & 127 & 7 & 5 & 128 & 22 & \color{red}1045 & 16 & 60 & \textbf{1508} \\
\bottomrule %
\end{tabular}
\end{table}

\newcommand{\customplot}[2]{
\begin{tikzpicture}
\begin{axis}
[
small,
width=0.35\textwidth,
height=5.5cm,
legend pos=south east,
ylabel=Average #2,
y label style={at={(axis description cs:.17,.5)},anchor=south,font=\small},
xmin=0,
xmajorgrids,
ymajorgrids,
xtick={0,10,...,65},
legend entries={Carbohydrate, Calorie, Fat, Protein, Sugar},
legend style={font=\scriptsize},
]
\addplot[color=purple,ultra thick,dashed] table{figs/#1/carbo.txt};
\addplot[color=orange,ultra thick,dotted] table{figs/#1/calorie.txt};
\addplot[color=darkgray,ultra thick,dashdotted] table{figs/#1/fat.txt};
\addplot[color=violet,ultra thick,dotted] table{figs/#1/protein.txt};
\addplot[color=cyan,very thick,] table{figs/#1/sugar.txt};
\end{axis}
\end{tikzpicture}
}

\begin{figure*}
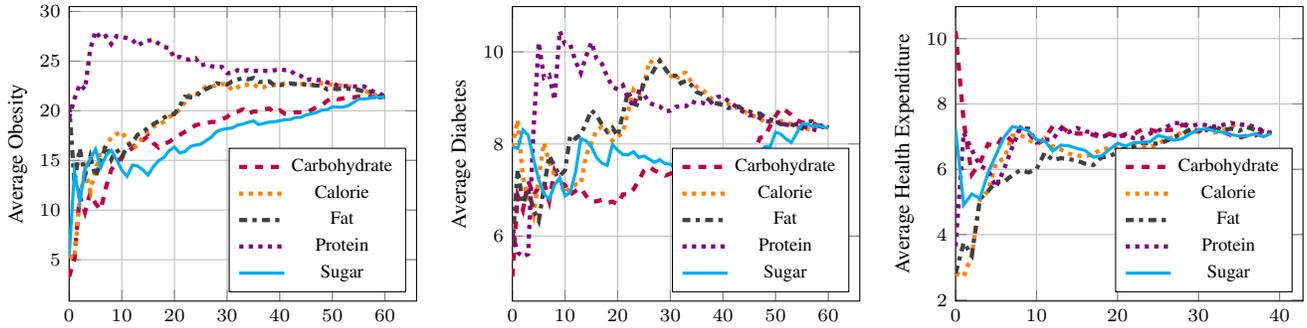

\centering
\subfloat{
	\customplot{obesity}{Obesity}
	\label{fig:nutitions-health:obesity}}
\hfil
\subfloat{
	\customplot{diabetes}{Diabetes}
	\label{fig:nutitions-health:diabetes}}
\hfil
\subfloat{
\customplot{medical}{Health Expenditure}
\label{fig:nutitions-health:medical}}
\caption{Average health measures of bottom-$k$ countries, based on nutrition values in their recipes. The values on x-axis indicate different values of $k$. As the value of $k$ increases, the countries with higher amounts of nutrition values contribute to the average.}
\label{fig:nutitions-health}
\end{figure*}

\section{Health and Nutrition}\label{sec:nutrition}
In this section, we investigate the relation between the nutrition values of the recipes associated with countries and their hard measures of health, including obesity rate, diabetes rate and health expenditure. Similar to Section \ref{sec:ingredients}, we map cuisines to countries by assigning all the recipes of those cuisines that relates to a specific country. Afterwards, we calculate the average calorie, protein, fat, carbohydrate and sugar values for each country over its recipes, weighted by user provided ratings as a measure of recipe popularity. Then, we calculated the correlations between average nutrition values and health measures. Pearson correlation, which captures the linear correlation between the two variables, and Kendall-Tau correlation, which measures the ordinal correlation, have been used for this task. The result is presented in Table \ref{table:results}. As the results suggest, nutrition values show a significant correlation with the health related measures of countries. The dominant positively correlated nutrients are the sugar and carbohydrate. It is intuitive because those are the main elements of snack meals like cakes, creams, etc which can contribute to the health difficulties and the consequence expenditures eventually. On the other hand, protein value shows strong negative correlation with the level of obesity and diabetes in countries. Noticeably, the positive impact of high-protein diets on losing weight is frequently studied in the literature \cite{halton2004effects}. 

Figure \ref{fig:nutitions-health} exhibits the relationship between the nutrients and health measures from a different perspective. In Figure \ref{fig:nutitions-health:obesity}, the average obesity of the $k$ countries intaking the least amounts of different nutrients (shown in different colors and line-styles) is plotted against the value of $k$. The same is shown for diabetes and health expenditure in Figures \ref{fig:nutitions-health:diabetes} and \ref{fig:nutitions-health:medical}, respectively. The trend of the diagrams endorses that including the countries with higher average nutrition values (except protein) results in an increase in the average health measures (e.g. average obesity). Proteins show completely opposite patterns as expected. Including the countries with higher protein diets decreases the rate of health difficulties (e.g. obesity or diabetes).
A noticeable trait in both Table \ref{table:results} and 
Figure \ref{fig:nutitions-health} is that the correlations and trends are more highlighted in the obesity results rather than the diabetes and health expenditure. The reason is that the diabetes and health expenditure are more elaborate phenomena than the obesity. For example, in addition to consuming foods, there are a variety of other genetic and environmental factors that may cause the diabetes. Remarkably, the genetic susceptibility of different ethnics varies so much \cite{elbein2009genetics}. As another example, over intaking of proteins itself can lead to an spectrum of adverse effects \cite{delimaris2013adverse}. Therefore the relation of protein intaking and health expenditures of the countries is not as clear as the relation between obesity and proteins.

\begin{table}
\small
\definecolor{tablegreen}{rgb}{1.0, 0.03, 0.0}
\definecolor{tablered}{rgb}{0, 0.54,0.27 }
\centering
\caption{Correlation of Different Health Measures with Nutrition Values of Recipes}
\label{table:results}
\begin{tabular}{l l r r}
\toprule
& &\multicolumn{2}{c}{Correlation Values} \\
\cmidrule(l){3-4}
Health Measure & Nutrient & Pearson & Kendall-Tao\\
\midrule
\multirow{5}{*}{Obesity} & Calorie  & $-0.104$ & $-0.110$ \\
& Protein  & {\color{tablered}\bm{$-0.483$}} & {\color{tablered}\bm{$-0.299$}} \\
& Fat  & $-0.115$ & $-0.127$ \\
& Carbohydrate  & $0.300$ & $0.201$ \\
& Sugar  & {\color{tablegreen}$\bm{0.461}$} & {\color{tablegreen}$\bm{0.293}$}  \\

\cmidrule(l){1-4}
\multirow{5}{*}{Diabetes} & Calorie & $-0.077$ & $-0.048$\\
& Protein  & {\color{tablered}\bm{$-0.162$}} & $-0.022$ \\
& Fat & $-0.123$ & {\color{tablered}\bm{$-0.063$}}  \\
& Carbohydrate  & {\color{tablegreen}$\bm{0.173}$} & {\color{tablegreen}\bm{$0.106$}} \\
& Sugar & $0.142$ & $0.066$ \\

\cmidrule(l){1-4}
\multirow{5}{*}{Health Expend.} & Calorie & $0.098$ & $0.110$\\
& Protein  & {\color{tablered}\bm{$-0.083$}} & {\color{tablered}\bm{$-0.022$}} \\
& Fat & {\color{tablegreen}$\bm{0.197}$} & {\color{tablegreen}$\bm{0.141}$}  \\
& Carbohydrate  & $-0.064$ & $-0.015$ \\
& Sugar & $0.134$ & $0.069$ \\
\bottomrule
\end{tabular}
\end{table}

\section{Related Work}\label{sec:related}

Recently, public health has been increasingly analyzed through the lens of the web and social media. We refer the reader to \cite{capurro2014use} for an overview of the recent research in this area. Abbar et al.~\cite{abbar2015you} relate food mentions on Twitter conversations to the obesity and diabetes rates, using caloric values, and find a high correlation (coefficient 0.77) between caloric values of tweets and obesity values in various states in the US. Low-obesity areas of USA have also been shown to be more socially active on Instagram (posting comments and likes) than those from high-obesity ones by Mejova et al.~\cite{mejova2015foodporn}, who present a large-scale analysis of pictures taken at 164K restaurants in the US. Silva et al.~\cite{Silva14:youare} identify cultural boundaries and similarities across populations at different scales based on the analysis of Foursquare check-ins.

Ahn et al.~\cite{ahn2011flavor} study culture-specific ingredient connections, creating a ``flavor network'' from a dataset of about 56K recipes and relating them to the geographical groupings of countries. Similar ``flavor-based'' food pairing studies are conducted on cuisines in distinct geographical areas such as India \cite{jain2015analysis}. West et al.~\cite{west2013cookies} mine logs of recipe-related queries to uncover temporal patterns in consumption. Using Fourier transforms, they show the yearly and weekly periodicity in food ``density'' of the searched recipes, with different trends in Southern and Northern hemispheres, suggesting a link between food selection and climate. A study of Austrian recipe sites by Wagner et al.~\cite{Wagner:2014:STP:2567948.2576951} also highlights differences in the recipes of regions which are further apart. Zhu et al.~\cite{zhu2013geography} conduct a similar study on Chinese recipes to investigate the effect of geographical and climatic proximities on ingredients similarity of domestic cuisines.

Kular et al.~\cite{kular2011using} create a network of recipes using a dataset of 300 recipes from 15 different countries, and show the network's {\em small-world} and scale-free properties. As opposed to this line of work, we also exploit  flavor and nutritional information, alongside health statistics countries to provide a deeper analysis about the dishes, cuisines, culinary cultures, and the impact of food on human life. Su et al.~\cite{su2014automatic} investigate underlying connections between cuisines and ingredients via machine learning classification, with an application to predicting the cuisine by looking at recipes. Like our analysis, theirs is based on a large-scale data collection of recipes---specifically, 226K recipes collected from food.com. However, they only look at classifying cuisines using Support Vector Machine (SVM), while we propose a deep neural network architecture to capture the highly non-linear relation of a recipe cuisine and its associated ingredients. The results approve that the proposed deep model outperforms SVM by a significant margin in terms of prediction accuracy and F-measure. 

There are major differences between our work and the ones discussed above, in both scale and domain. An important characteristic of our work comes from the size and the quality of the various datasets we used, which enable us to derive first-of-its-kind insight on worldwide cuisines and their relationship to health factors. In addition to the ingredients, we also exploited flavor and nutritional information, alongside health and immigration statistics, allowing us to perform a deeper analysis of the dishes, cuisines, culinary cultures, as well as the impact of food on human life.

\section{Conclusion}
\label{sec:conclusions}

This paper presented a large-scale study of user-generated recipes on the web, their ingredients, nutrition, similarities across countries, and their relation with country health statistics. Our results have multiple implications: we found strong similarities between cuisines in neighboring countries, yet, the diversity of ingredients and flavors varies largely across the continents, mostly affected by net migration trends. %
We found quantitative evidence of a strong correlation between nutrition information of the recipes (e.g., in terms of sugar intake) and obesity. Also, we demonstrated that deep learning can be used to effectively  predicting cuisines from ingredients, potentially providing possibility for fine-grained analysis of food and dishes as well as improved recipe recommendations based on individuals' profile.

Our findings indicate that certain ingredients (e.g., mozzarella) uniquely represent a certain cuisine (e.g., Italian) and there are strong clusters of ingredients across neighboring countries. This feature eases the prediction of regions (e.g., continents) from the combination of ingredients in a cuisine. Moreover, the correlation between ingredients and health conditions, such as diabetes, can be very useful to public health experts, where behavior nudges or recommendation of similar dishes in flavor and ingredient complexity can be utilized to improve dietary intake~\cite{regulating2011judging, foster2005behavioral}.

In future work, we plan to explore the possibility of recipe recommendation based on regional and personal tastes and user ratings. This is important as a local Chinese dish or a distinct flavor combination may be ``alien'' to, e.g., a Western person, but of interest to a Japanese individual. We also wish to asses the ability to model flavors with ingredients, and discover ingredients to match a specific flavor palette. Finding answers to these questions would provide a better understanding of the composition of flavors and ingredients in popular dishes and provide a better recommendation system for a healthier, tastier, and more diverse experience.

\bibliographystyle{abbrv}

\end{document}